\documentstyle[preprint,prd,aps]{revtex}

\begin{document}
\draft
\preprint{\ \vbox{
\halign{&##\hfil\cr
       AS-ITP-2000-03\cr\cr}} } \vfil

\title{Relativistic Corrections for Polarized $J/\Psi$-Production in b-decay}
\author{J.P. Ma}
\address{Institute of Theoretical Physics,\\
Academia Sinica, \\
P.O.Box 2735, Beijing 100080, China\\
e-mail: majp@itp.ac.cn}
\maketitle

\begin{abstract}
We analyse the structure of relativistic corrections in the inclusive
production of polarized $J/\Psi$ from b-quark decay. The analysis is
performed not only for the production channel in which the $c\bar c$ pair is
a color sinlet, but also for the channels in which the $c\bar c$ is a color
octet. We find that the correction in the color-singlet channel at the
tree-level is completely determined by the decay constant of $J/\Psi$, while
in the color-octet channels the corrections are characterized by three
matrix elements defined in NRQCD, whose numerical values are unknown. We
discuss the impact of these corrections on the polarized $J/\Psi$%
-production, and the impact is so significant that the predictions based on
the analysis for the considered process may be unreliable. Finally, we
propose an integrated spin observable to measure the polarization of $J/\Psi$%
. \newline
\end{abstract}

\preprint{\ \vbox{
\halign{&##\hfil\cr
       AS-ITP-2000-03\cr\cr}} } \vfil

\eject
\baselineskip=20pt

\vspace{-5mm}

\vskip20pt \narrowtext
Quarkonium systems generally are thought as simpler than light hadrons and
it may be easier to handle them in the framework of QCD. However, only
recently we have been able to treat their decays and productions rigorously,
based on a factorization with non-relativistic QCD(NRQCD)\cite{BBL}. In the 
framework of the NRQCD factorization 
the effect of short distance is handled with perturbative QCD and the effect
of long distance is parameterized with NRQCD matrix elements. The
factorization is performed by utilizing the fact that a heavy quark inside a
quarkonium moves with a small velocity $v$ in the quarkonium rest frame and
an expansion in $v$ can be employed. In this factorization, inclusive
productions of a quarkonium, e.g., like $J/\Psi $, can be imagined at the
leading order of $v$ as the following: The $c$- and $\bar{c}$- quark are
produced at a same space-time point, and then this pair is transformed into
the $J/\Psi $. The production of the $c\bar{c}$ pair can be calculated with
perturbative QCD, while the transformation is a nonperturbative process,
which can be described by matrix elements defined in NRQCD. 
At higher order of $v$ various effects
are taken into account, e.g., relativistic effect,
and the effect of the produced $c\bar{c}$ pair which does not have the
same quantum numbers as those of $J/\Psi $. For $J/\Psi $ the transition of
a $c\bar{c}$ pair in a color-octet state happens at higher orders in the
small velocity expansion, but in some cases, the produced $c\bar{c}$ pair is
more likely in a color-octet state than in a color-singlet state, hence the
production of $J/\Psi $ through a color-octet $c\bar{c}$ pair is not
negligible, and has an important contribution to the total production rate.
Including this contribution one is able to fitting experimental results of
the total production rate obtained in Tevatron\cite{BF,CL,BK}.

\vskip20pt With the NRQCD factorization, not only the total production rate
of a quakonium can be predicted, but also the polarization of the
quarkonium, if it has a spin. Recent preliminary measurement\cite
{CDF} at Tevatron by CDF shows that the produced $J/\Psi $ is polarized in
the way which is unexpected from theoretical predictions. Several attempts
to explain the discrepancy are made\cite{BKL,JM}. It should be kept in mind
that these predictions are only based the theoretical analysison 
at the leading order in $\alpha _s$ and at the leading order in $v$. 
For charmonia the velocity is not very small, typically $%
v^2\approx 0.3$, and one-loop effects are generally substantial. Adding effects
at higher orders of $\alpha _s$ and of $v$, the predictions may be changed
significantly. In this work we study the effects of higher orders in $v$ on
the polarization of $J/\Psi $ produced in b-quark decay, to see how
important the effects are.

\vskip20pt In the framework of heavy quark effective theory inclusive decays
of a b-flavored hadron can be thought approximately as inclusive decays of a
free b-quark. The inclusive $b$-quark decay into a $J/\Psi $ were studied at
the leading order in $v$ in \cite{KLS,FH}, in \cite{BMR} the one-loop
correction from QCD for the unpolarized decay was studied. In these studies
it is already shown that the color-octet $c\bar{c}$ pair plays an important
role. In this work we will not only considered the color-singlet
contribution but also the color-octet contribution. With the NRQCD
factorization, the polarized decay width for the process 
\begin{equation}
b\rightarrow J/\Psi +X
\end{equation}
can be written as: 
\begin{equation}
\Gamma _\lambda (J/\Psi )=\sum_nC_n(b\rightarrow c\bar{c}[n]+X)\langle
0|O^{J/\Psi }[n]|0\rangle ,
\end{equation}
in which the coefficients $C_n(b\rightarrow c\bar{c}[n]+X)$ describe the
production of a $c\bar{c}$ pair in a state $n$, the matrix elements $\langle
0|O^{J/\Psi }[n]|0\rangle $ characterize the transition of a $c\bar{c}$ pair
in $n$-state into the $J/\Psi $. The coefficients can be calculated with
perturbative QCD because the production of a $c\bar{c}$ pair is a
short-distance process, while the matrix elements represent nonperturbative
effects and they are defined with operators in NRQCD. These matrix elements
are scaled by the power of $v$ with the rule of power counting in $v$\cite
{LMN}. The index $\lambda $ stands for the helicity of the $J/\Psi $, $%
\lambda =L$ is for the longitudinal polarization, $\lambda =T$ is for the
transversal polarization. We will use the matching procedure proposed in 
\cite{BBL} to identify the operators at the next-to-leading order in $v$ and
to calculate the corresponding coefficients.

\vskip20pt To do the matching we need to consider the process 
\begin{equation}
b(p_b)\rightarrow c(p_1)+\bar{c}_{(}p_2)+X,
\end{equation}
where the momenta are given in the brackets. With a Lorentz transformation
we can boost the $c\bar{c}$ pair into its rest-frame and denote in the
rest-frame the three- momentum of $c$ and of $\bar{c}$ as ${\bf {q}}$ and $-%
{\bf {q}}$ respectively. The total decay width for the process in Eq.(3) can
be written 
\begin{equation}
\Gamma (b\rightarrow c\bar{c}+X)=\int \frac{d^3q}{(2\pi )^3}\hat{\Gamma}(%
{\bf q)}
\end{equation}
The matching condition reads: 
\begin{equation}
\hat{\Gamma}({\bf {q}})=\sum_nC_n(b\rightarrow c\bar{c}[n]+X)\langle 0|O^{c%
\bar{c}}[n]|0\rangle .
\end{equation}
In the above equation, the ${\bf {q}}$-dependence in the right hand side is
only contained in the matrix elements, in which the hadronic state is
replaced by the partonic $c\bar{c}$ state. With Eq.(5) one can identify the
matrix elements appearing in Eq.(2) and can determine the corresponding
coefficients.

\vskip20pt 
The effective weak Hamiltonian for $b$-quark decay is: 
\begin{eqnarray}
H_{{\rm eff}} &=&\frac{G_F}{\sqrt{2}}\sum_{q=s,d}\big\{V_{cb}V_{cq}^{*}[\frac 13%
C_1(\mu )\bar{c}\gamma ^\mu (1-\gamma _5)c\bar{q}\gamma _\mu (1-\gamma _5)b 
\nonumber \\
&\ &+C_8(\mu )\bar{c}T^a\gamma ^\mu (1-\gamma _5)c\bar{q}T^a\gamma _\mu
(1-\gamma _5)b]\big\}.
\end{eqnarray}
We neglected the contributions of QCD penguin operators in $H_{{\rm eff}}$. $%
T^a$($a=1,\cdots 8$) is SU(3) color-matrix. The coefficients $C_1$ and $C_8$
are related to the usual $C_{\pm }$ by 
\begin{equation}
C_1(\mu )=2C_{+}(\mu )-C_{-}(\mu ),\ \ \ C_8(\mu )=C_{+}(\mu )+C_{-}(\mu ).
\end{equation}
With the one-loop evolution one can obtain 
\begin{equation}
\frac{C_1(m_b)}{C_8(m_b)}\approx 0.18
\end{equation}
This indicates that the $c\bar{c}$ pair is produced more likely in a
color-octet state than in a color-singlet state. Hence the color-octet
contribution is significant and should be included in the decay width. With
the effective Hamiltonia it is straightforward to calculate $\hat{\Gamma}(%
{\bf q)}$ and to determine the coefficients in Eq.(5). However, the power
counting of $v$ for matrix elements with partonic states is different than
that for matrix elements with hadronic states. Before presenting our
results, we briefly discuss the identification of operators in Eq.(2). In
general the matrix elements in Eq.(2) take the form 
\begin{equation}
\langle 0|O^H[n]|0\rangle =\sum_X\langle 0|{\cal K}_n|H+X\rangle \langle H+X|%
{\cal K}_n^{\prime }|0\rangle ,
\end{equation}
where $H$ denotes a quarkonium. ${\cal K}_n$ and ${\cal K}_n^{\prime }$ are operators
defined in NRQCD. The general constraints can be obtained by considering
symmetries of QCD or NRQCD. Because QCD respects charge conjugation- and
parity symmetry, the operators must be transformed in the same way under C-
and P transformation. The operators ${\cal K}_n$ and ${\cal K}_n^{\prime }$
can be classified with the weight $j$ of $SU(2)$ representations. Since the
total angle momentum is conserved, the operators must have the same weight.
By considering the approximated symmetry, spin-symmetry of NRQCD, further
constraints can be obtained. With these constraints, one can identify the
operators appearing in Eq.(2).

\vskip20pt Now we give our results for $\Gamma _\lambda (J/\Psi )$. We write 
\begin{equation}
\Gamma _\lambda (J/\Psi )=\Gamma _\lambda ^{(1)}+\Gamma _\lambda ^{(8)},
\end{equation}
where the index 1 or 8 stands for color singlet contributions or for
color-octet contributions, respectively. The leading order in $v$ for color
singlet contributions is $v^0$, while the leading order for color-octet
contributions is $v^4$. We expand the quantities: 
\begin{eqnarray}
\Gamma _\lambda ^{(1)} &=&\Gamma _\lambda ^{(1,v^0)}+\Gamma ^{(1,v^2)}+{\cal %
O}(v^4),  \nonumber \\
\Gamma _\lambda ^{(8)} &=&\Gamma _\lambda ^{(8,v^4)}+\Gamma ^{(8,v^6)}+{\cal %
O}(v^8).
\end{eqnarray}
For color-singlet contributions at the leading order, there is only one
matrix element representing the transformation of a $c\bar{c}$ pair in $%
^3S_1 $ state into the $J/\Psi $. At this order we have: 
\begin{eqnarray}
\Gamma _T^{(1,v^0)} &=&\frac{C_1^2G_F^2|V_{cb}|^2}{432\pi m_b^3m_c}%
(m_b^2-4m_c^2)^2\langle 0|O_1^{J/\Psi }(^3S_1)|0\rangle \cdot 4m_c^2, 
\nonumber \\
\Gamma _L^{(1,v^0)} &=&\frac{C_1^2G_F^2|V_{cb}|^2}{432\pi m_b^3m_c}%
(m_b^2-4m_c^2)^2\langle 0|O_1^{J/\Psi }(^3S_1)|0\rangle \cdot m_b^2,
\end{eqnarray}
where we used $|V_{cd}|^2+|V_{cs}|^2\approx 1$. At order of $v^2$, there is
also one matrix element which represents the effect of the relative movement
of the $c\bar{c}$ pair inside $J/\Psi $. The results reads: 
\begin{equation}
\Gamma _\lambda ^{(1,v^2)}=\frac{C_1^2G_F^2|V_{cb}^2}{432\pi m_c}%
m_b(m_b^2-4m_c^2)\frac{\langle 0|P_1^{J/\Psi }(^3S_1)|0\rangle }{m_c^2}\cdot
G_\lambda (\frac{m_c^2}{m_b^2})
\end{equation}
with 
\begin{eqnarray}
G_L(y) &=&-\frac{3+92y+208y^2+576y^3}{6(1+4y)^2},  \nonumber \\
G_T(y) &=&\frac{2y(3-68y-304y^2-960y^3)}{3(1+4y)^2}.
\end{eqnarray}
The definition of the matrix elements $\langle 0|O_1^{J/\Psi
}(^3S_1)|0\rangle $ and $\langle 0|P_1^{J/\Psi }(^3S_1)|0\rangle $ can be
found in \cite{BBL}. These matrix elements can only be calculated
nonperturbatively and they must be calculated with the same accuracy in
order of $v$ if one uses both to make predictions.

\vskip20pt 
For color octet contributions at the leading order in $v$, the $c%
\bar{c}$ pair can be in $^1S_0$, $^3S_1$ and $^3P_1$ states, these states
can be transformed into a $J/\Psi $ through emission or absorption of soft
gluons. Correspondingly, we have in the matching three types of matrix
elements, these matrix elements can be reduced to three matrix elements with
the symmetries mentioned before. They are: 
\begin{eqnarray}
&&\sum_X\langle 0|\chi ^{\dagger }T^a\psi |J/\Psi +X\rangle \langle J/\Psi
+X|\psi ^{\dagger }T^a\chi |0\rangle   \nonumber \\
&&\ \ \ =\frac 13\varepsilon _i(\lambda )\varepsilon _i^{*}(\lambda )\langle
0|O_8^{J/\Psi }(^1S_0)|0\rangle  \\
&&\sum_X\langle 0|\chi ^{\dagger }T^a\sigma ^i\psi |J/\Psi +X\rangle \langle
J/\Psi +X|\psi ^{\dagger }T^a\sigma ^j\chi |0\rangle   \nonumber \\
&&\ \ \ =\frac 13\varepsilon _i(\lambda )\varepsilon _j^{*}(\lambda )\langle
0|O_8^{J/\Psi }(^3S_1)|0\rangle  \\
&&\sum_X\langle 0|\chi ^{\dagger }T^a({\bf \sigma }\times (-\frac i2%
{\tensor{\bf {D}}}
))^i\psi |J/\Psi +X\rangle \langle J/\Psi +X|\psi ^{\dagger }T^a{\bf \sigma }%
\times (-\frac i2\tensor{\bf D}))^j\chi |0\rangle   \nonumber \\
&&\ \ \ =\frac 13(\delta _{ij}{\bf \varepsilon }(\lambda )\cdot {\bf %
\varepsilon }^{*}(\lambda )-\varepsilon _i(\lambda )\varepsilon
_j^{*}(\lambda )\langle 0|O_8^{J/\Psi }(^3P_1)|0\rangle 
\end{eqnarray}
In the above equations $\varepsilon _i(\lambda )(i=1,2,3)$ denotes the
polarization vector of $J/\Psi $ in its rest-frame. The field $\psi (\chi )$
is the field in NRQCD for the $c(\bar{c})$ quark. ${\bf D}$ is the spacial
part of the covariant derivative $D^\mu $. The definition of the operator $%
O_8^{J/\Psi }(^1S_0)$, $O_8^{J/\Psi }(^3S_1)$ and $O_8^{J/\Psi }(^3P_1)$ can
be found in \cite{BBL}. The rotation invariance leads to Eq.(15), while spin
symmetry is used for Eq.(16) and Eq.(17). The leading order of these
matrices is at $v^4$. It should be noted that Eq.(16) and Eq.(17) also hold
at order of $v^6$, because to keep the correct spin configuration the spin
of the $c\bar{c}$ pair must be flipped twice through the interaction which
violates the spin symmetry, this can only happen at order of $v^8$. With the
above identification we obtain the color-octet contribution at the leading
order: 
\begin{eqnarray}
\Gamma _L^{(8,v^4)} &=&\frac{C_8^2G_F^2|V_{cb}|^2}{288\pi m_b^3m_c}%
(m_b^2-4m_c^2)^2  \nonumber \\
&\cdot &\left\{ m_b^2\langle 0|O_8^{J/\Psi }(^1S_0)|0\rangle +m_b^2\langle
0|O_8^{J/\Psi }(^3S_1)|0\rangle +{8}\langle 0|O_8^{J/\Psi }(^3P_1)|0\rangle
\right\}   \nonumber \\
\Gamma _T^{(8,v^4)} &=&\frac{C_8^2G_F^2|V_{cb}|^2}{288\pi m_b^3m_c}%
(m_b^2-4m_c^2)^2  \nonumber \\
&\cdot &\left\{ m_b^2\langle 0|O_8^{J/\Psi }(^1S_0)|0\rangle +4m_c^2\langle
0|O_8^{J/\Psi }(^3S_1)|0\rangle +\frac{m_b^2+4m_c^2}{m_c^2}\langle
0|O_8^{J/\Psi }(^3P_1)|0\rangle \right\} .
\end{eqnarray}

\vskip20pt 
At order of $v^6$ for the color-octet contributions various
matrix elements appear at the first look. For example, the matrix element 
\begin{equation}
\sum_X\langle 0|\chi ^{\dagger }T^a\sigma ^l(-\frac i2)^2{{%
{\tensor{{\bf D}}}
} }^{(l}{{%
{\tensor{{\bf D}}}
} }^{i)}\psi |J/\Psi +X\rangle \langle J/\Psi +X|\psi ^{\dagger }T^a\sigma
^j\chi |0\rangle +h.c..
\end{equation}
To take a close look at the matrix element we consider 
\begin{equation}
B^{ijlk}=\sum_X\langle 0|\chi ^{\dagger }T^a\sigma ^i(-\frac i2)^2{{%
{\tensor{{\bf D}}}
} }^{(l}{{%
{\tensor{{\bf D}}}
} }^{k)}\psi |J/\Psi +X\rangle \langle J/\Psi +X|\psi ^{\dagger }T^a\sigma
^j\chi |0\rangle +h.c..
\end{equation}
At order of $v^6$ the spin symmetry still can be used for the matrix
element, it leads: 
\begin{equation}
B^{ijlk}=\varepsilon _i(\lambda )\varepsilon _j^{*}(\lambda )\cdot A_{lk}.
\end{equation}
By rotation invariance, $A_{lk}$ is proportional to $\delta _{lk}$. Because
the tensor ${%
{\tensor{{\bf D}}}
} ^{(l}{%
{\tensor{{\bf D}}}
} ^{k)}$ is symmetric and trace-less, we conclude that $B^{ijlk}$ is zero at
this order, hence also the matrix element in Eq.(19). This can also be
understood as the following: Because the spin symmetry holds, the total
orbit angle moment is conserved. The $X$-state in the bra in Eq.(20) is a $%
l=0$ state because of the conservation, while the $X$-state in the ket is a $%
l=2$ state. Therefore, the matrix element is excluded by the definition of
the sum over $X$ state. By considering all constraints, only three types of
matrix elements remain, and they can be further reduced to three matrix
elements: 
\begin{eqnarray}
&&\sum_X\langle 0|\chi ^{\dagger }T^a(-\frac i2{{%
{\tensor{{\bf D}}}
} })^2\psi |J/\Psi +X\rangle \langle J/\Psi +X|\psi ^{\dagger }T^a\chi
|0\rangle +h.c.  \nonumber \\
&&\ \ \ =\frac 23\varepsilon _i(\lambda )\varepsilon _i^{*}(\lambda )\langle
0|P_8^{J/\Psi }(^1S_0)|0\rangle \\
&&\sum_X\langle 0|\chi ^{\dagger }T^a\sigma ^i(-\frac i2{{%
{\tensor{{\bf D}}}
} })^2\psi |J/\Psi +X\rangle \langle J/\Psi +X|\psi ^{\dagger }T^a\sigma
^j\chi |0\rangle +h.c.  \nonumber \\
&&\ \ \ =\frac 23\varepsilon _i(\lambda )\varepsilon _j^{*}(\lambda )\langle
0|P_8^{J/\Psi }(^3S_1)|0\rangle \\
&&\sum_X\langle 0|\chi ^{\dagger }T^a({\bf \sigma }\times (-\frac i2{%
{\tensor{{\bf D}}}
} ))^i(-\frac i2{%
{\tensor{{\bf D}}}
} )^2\psi |J/\Psi +X\rangle \langle J/\Psi +X|\psi ^{\dagger }T^a{\bf \sigma 
}\times (-\frac i2{%
{\tensor{{\bf D}}}
} ))^j\chi |0\rangle +h.c.  \nonumber \\
&&\ \ \ =\frac 23(\delta _{ij}{\bf \varepsilon }(\lambda )\cdot {\bf %
\varepsilon }^{*}(\lambda )-\varepsilon _i(\lambda )\varepsilon
_j^{*}(\lambda )\langle 0|P_8^{J/\Psi }(^3P_1)|0\rangle .
\end{eqnarray}
The matrix elements $\langle 0|P_8^{J/\Psi }(^1S_0)|0\rangle $, $\langle
0|P_8^{J/\Psi }(^3S_1)|0\rangle $ and $\langle 0|P_8^{J/\Psi
}(^3P_1)|0\rangle $ can be obtained by summing over the helicity and by the
contraction over the indices in the above equations. With these matrix
elements the results for the color-octet contribution at order of $v^6$ are: 
\begin{eqnarray}
\Gamma _\lambda ^{(8,v^6)} &=&\frac{C_8^2G_F^2|V_{cb}|^2}{288\pi m_c^3}%
m_b(m_b^2-4m_c^2)  \nonumber \\
&\cdot &\big\{ F(\frac{m_c^2}{m_b^2})\cdot \langle 0|P_8^{J/\Psi
}(^1S_0)|0\rangle +G_\lambda (\frac{m_c^2}{m_b^2})\cdot \langle
0|P_8^{J/\Psi }(^3S_1)|0\rangle  \nonumber \\
&&\ \ +H_\lambda (\frac{m_c^2}{m_b^2})\cdot \frac 1{m_c^2}\langle
0|P_8^{J/\Psi }(^3P_1)|0\rangle \big\}
\end{eqnarray}
with 
\begin{eqnarray}
F(y) &=&-\frac{7+108y+144y^2+320y^3}{6(1+4y)^2},  \nonumber \\
H_L(y) &=&-\frac{4y(1+84y+240y^2+704y^3)}{6(1+4y)^2},  \nonumber \\
H_T(y) &=&-\frac{7+84y+144y^2+704y^3}{6(1+4y)^2}.
\end{eqnarray}
The function $G_\lambda (y)$ is given in Eq.(14). The above results indicate
that the correction in the color-octet contributions is to take the effect
of the relative movement of the $c\bar{c}$ pair inside the $J/\Psi $ into
account. This is similar as the case with the color-singlet contribution.
Hence, the total corrections are just relativistic corrections characterized
by four matrix elements.

\vskip20pt With the results given above one may predict the polarization of
the produced $J/\Psi $ and the total decay width $\Gamma =2\Gamma _T+\Gamma
_L$, if one knows the numerical values of the eight matrix elements. Among
the four matrix elements at the leading order in $v$ the best known matrix
element is $\langle 0|O_1^{J/\Psi }(^3S_1)|0\rangle $, which is calculated
with potential models and with lattice QCD, while the other three are
extracted from experimental data. The value of $\langle 0|O_8^{J/\Psi
}(^3S_1)|0\rangle $ is rather well determined by direct $J/\Psi $ production
at large transverse momentum in $p\bar{p}$ collisions\cite{BF,CL,BK}.
However the uncertainty for this determination is large and it can be at the
level of $100\%$. The other two are not well determined, only certain
combination of them is know with a large uncertainty. In \cite{KK} 
a re-analysis of experimental data from Tevatron and from Hera is performed, 
in which some higher-orders effects due to multiple-gluon
initial-states radiation are considered. It is shown that the values
of these matrix elements can be changed substantially by including these
effects. Based on the leading
order results an analysis for predictions of the polarized $J/\Psi $ is given%
\cite{FH}. The four matrix elements at the next-to-leading order in $v$ are
completely unknown. With the power-counting in $v$, we only know that they
are suppressed by $v^2$ relatively to the corresponding matrix elements at
the leading order in $v$. All of these prevents us from a detailed
prediction for the polarization and for the total decay width. Nevertheless,
one can still see the impact of these corrections. For the color-singlet
contributions, if one neglects the one-loop QCD correction and uses the
vacuum saturation, then with $H_{{\rm eff}}$ in Eq.(6) one has only one
constant representing the nonperturbative effect in the process in Eq.(1): 
\begin{equation}
\langle J/\Psi |\bar{c}\gamma ^\mu c|0\rangle =-if_{J/\Psi }M_{J/\Psi
}(\varepsilon ^\mu (\lambda ))^{*},
\end{equation}
where $f_{J/\Psi}$ is the leptonic decay constant of $J/\Psi $ and 
is related to the
leptonic decay width: 
\begin{equation}
\Gamma (J/\Psi \rightarrow \ell ^{+}\ell ^{-})=\frac{16\pi \alpha _{{\rm em}%
}^2}{27M_{J/\Psi }}f_{J/\Psi }^2.
\end{equation}
In the above equation, the only approximation is to neglect effects of
higher orders in $\alpha _{{\rm em}}$ and lepton masses, relativistic
corrections are automatically included. In NRQCD the vacuum saturation
brings uncertainty at order of $v^4$. Therefore the color-singlet
contributions can be written as: 
\begin{eqnarray}
\Gamma _T^{(1)} &=&\frac{C_1^2G_F^2|V_{cb}|^2}{144\pi m_b^3}(m_b^2-M_{J/\Psi
}^2)^2M_{J/\Psi }^2f_{J/\Psi }^2+{\cal O}(\alpha _s^2)+{\cal O}(v^4), 
\nonumber \\
\Gamma _L^{(1)} &=&\frac{C_1^2G_F^2|V_{cb}|^2}{144\pi m_b^3}(m_b^2-M_{J/\Psi
}^2)^2m_b^2f_{J/\Psi }^2+{\cal O}(\alpha _s^2)+{\cal O}(v^4),
\end{eqnarray}
where the relativistic correction calculated before is contained in the $%
J/\Psi $-mass $M_{J/\Psi }$ and in the decay constant. Comparing the results
in Eq.(29) with that in Eq.(12) we may see how large the relativistic
correction is. For this we define: 
\begin{equation}
R_\lambda ^{(1)}=\frac{\Gamma _\lambda ^{(1,v^0)}}{\Gamma _\lambda ^{(1)}},
\end{equation}
where $\Gamma _\lambda ^{(1,v^0)}$ is given in Eq.(12) and the results in
Eq.(29) are used for $\Gamma _\lambda ^{(1)}$. To obtain numerical values of 
$R_\lambda ^{(1)}$, we take $m_b=4.7$GeV and $m_c=1.5$GeV. $f_{J/\Psi }$ is
determined by the leptonic decay to be 405MeV. For the matrix element $%
\langle 0|O_1^{J/\Psi }(^3S_1)|0\rangle $, its value is determined based on
a potential model in \cite{EQ} to be 1.16(GeV$)^3$, which is in agreement
with a calculation of lattice QCD in \cite{BSK}. With these values we
obtain: 
\begin{equation}
R_L^{(1)}\approx R_T^{(1)}\approx 1.6.
\end{equation}
This indicates that the correction is negative and very large. Another way
to see the impact of the correction is to compare the coefficients in the
front of the matrix element. We take the above numerical values for quark
masses and obtain: 
\begin{eqnarray}
\Gamma _L^{(1)} &=&\frac{C_1^2G_F^2|V_{cb}|^2}{432\pi }\cdot \left\{
24.3\langle 0|O_1^{J/\Psi }(^3S_1)|0\rangle -52.4\frac 1{m_c^2}\langle
0|P_1^{J/\Psi }(^3S_1)|0\rangle \right\} +{\cal O}(v^4),  \nonumber \\
\Gamma _T^{(1)} &=&\frac{C_1^2G_F^2|V_{cb}|^2}{432\pi }\cdot \left\{
9.90\langle 0|O_1^{J/\Psi }(^3S_1)|0\rangle -11.4\frac 1{m_c^2}\langle
0|P_1^{J/\Psi }(^3S_1)|0\rangle \right\} +{\cal O}(v^4).
\end{eqnarray}
With these results the relativistic correction can be at the level of $60\%$
for $\Gamma _L^{(1)}$ and at the level of $35\%$ for $\Gamma _T^{(1)}$ if
one takes 
\begin{equation}
\frac 1{m_c^2}\langle 0|P_1^{J/\Psi }(^3S_1)|0\rangle \approx v^2\langle
0|O_1^{J/\Psi }(^3S_1)|0\rangle ,\ \ v^2\approx 0.3.
\end{equation}
Similarly we obtain for the color-octet contributions: 
\begin{eqnarray}
\Gamma _L^{(8)} &=&\frac{C_8^2G_F^2|V_{cb}|^2}{288\pi }\cdot \big\{ %
24.4\langle 0|O_8^{J/\Psi }(^1S_0)|0\rangle -68.3\cdot \frac 1{m_c^2}\langle
0|P_8^{J/\Psi }(^1S_0)|0\rangle  \nonumber \\
&&\ \ +24.4\langle 0|O_8^{J/\Psi }(^3S_1)|0\rangle -52.4\frac 1{m_c^2}%
\langle 0|P_8^{J/\Psi }(^3S_1)|0\rangle  \nonumber \\
&&\ \ +19.8\frac 1{m_c^2}\langle 0|O_8^{J/\Psi }(^3P_1)|0\rangle -35.8\frac 1%
{m_c^4}\langle 0|P_8^{J/\Psi }(^3P_1)|0\rangle \big\}+{\cal O}(v^8), 
\nonumber \\
\Gamma _T^{(8)} &=&\frac{C_8^2G_F^2|V_{cb}|^2}{288\pi }\cdot \big\{ %
24.4\langle 0|O_8^{J/\Psi }(^1S_0)|0\rangle -68.3\cdot \frac 1{m_c^2}\langle
0|P_8^{J/\Psi }(^1S_0)|0\rangle  \nonumber \\
&&\ \ +9.9\langle 0|O_8^{J/\Psi }(^3S_1)|0\rangle -11.4\frac 1{m_c^2}\langle
0|P_8^{J/\Psi }(^3S_1)|0\rangle  \nonumber \\
&&\ \ +34.3\frac 1{m_c^2}\langle 0|O_8^{J/\Psi }(^3P_1)|0\rangle -86.5\frac 1%
{m_c^4}\langle 0|P_8^{J/\Psi }(^3P_1)|0\rangle \big\}+{\cal O}(v^8).
\end{eqnarray}
If one assumes that there are similar relations among the color-octet matrix
elements like that in Eq.(33), then the correction is very large. For
example, the correction for the $^1S_0$ production channel can be at the level
of $80\%$, and the correction for the $^3P_1$ production channel can be at the
level of $70\%$. With the above discussions one may conclude that the
predictions based on the leading- and next-to-leading order in the small
velocity expansion for the process are unreliable.

\vskip20pt The last subject of this work is to propose an integrated
observable to measure the spin of the produced $J/\Psi $. Usually, one looks
at the leptonic decay of $J/\Psi $ to measure the spin. At a fixed momentum $%
{\bf P}$ of $J/\Psi $ one measures the distribution: 
\begin{equation}
\frac{d\Gamma }{d\cos \theta }(J/\Psi \rightarrow \ell ^{+}\ell ^{-})\propto
1+\alpha \cos ^2\theta ,
\end{equation}
where $\theta $ is the angle between ${\bf P}$ and ${\bf k}$, ${\bf k}$ is
the momentum of the lepton in the $J/\Psi $-rest frame. The parameter $%
\alpha $ is predicted as: 
\begin{equation}
\alpha =\frac{\Gamma _T-\Gamma _L}{\Gamma _T+\Gamma _L}.
\end{equation}
This may have a disadvantage that it will be hard to determine the
distribution if the number of events with a fixed ${\bf P}$ is small, hence
the parameter $\alpha $. We propose to use an integrated spin observable to
overcome this disadvantage. For this purpose we define the density matrix $%
R_{ij}$ of the produce $J/\Psi $. For arbitrary polarization the decay width
can be written: 
\begin{equation}
\Gamma (b\rightarrow J/\Psi +X)=\frac 1{4\pi }\int d\Omega \varepsilon
_iR_{ij}({\bf \hat{P}})\varepsilon _j,
\end{equation}
where $\Omega $ is the solid angle of ${\bf P}$, ${\bf \hat{P}}$ denotes the
direction of ${\bf P}$. Similarly we can define the density matrix $\rho
_{ij}$ for the leptonic decay. Any observable $O$ can be then predicted by 
\begin{equation}
\langle O\rangle =\frac 1N\int \frac{d\Omega }{4\pi }\int 
\frac{d\Omega _k}{4\pi }
O\cdot \rho _{ij}({\bf \hat{k}})\cdot R_{ji}({\bf \hat{P}}),
\end{equation}
where $N$ is a normalization factor so that $\langle 1\rangle =1$. A simple
calculation leads: 
\begin{equation}
\rho _{ij}({\bf \hat{k}})=\frac 13\delta _{ij}-\frac 12(\hat{k}_i\hat{k}_j-%
\frac 13\delta _{ij}),
\end{equation}
where $\rho _{ij}$ is normalized. $R_{ij}$ can be written: 
\begin{eqnarray}
R_{ij}({\bf \hat{P}}) &=&a\delta _{ij}+ib\varepsilon _{ijk}\hat{P}_k+c\hat{P}%
_i\hat{P}_j,  \nonumber \\
a &=&\Gamma _T,\ \ \ c=\Gamma _L-\Gamma _T.
\end{eqnarray}
The anti-symmetric part exits due to that parity is violated, and it is
irrelevant here because the parity is conserved in the leptonic decay. With
the tensor structure one can construct the integrated spin observable $O_P$,
and predict its value with Eq.(38): 
\begin{equation}
O_P=({\bf \hat{P}}\cdot {\bf \hat{k}})^2-\frac 13,\ \ \ \ \langle O_P\rangle
=\frac{2(\Gamma _T-\Gamma _L)}{15\Gamma }.
\end{equation}
If $\langle O_P\rangle =0$ the $J/\Psi $ is unpolarized. If one know the
values of the matrix elements discussed before, one can obtain the value for 
$\langle O_P\rangle $ to compare with the measured in experiment. One may
also obtain the invariance $\langle O_P^2\rangle $ to determine the
statistical error of $\langle O_P\rangle $.

\vskip20pt We summarize our work: We have analyzed the relativistic
corrections for the polarized $J/\Psi$-production in $b$-quark decay. We
have calculated the perturbative coefficients and identified the matrix
elements at the next-to-leading order in $v$. We find that the corrections
can be very large in the color-singlet production channel and as well as in
the color-octet production channels. For the color-singlet production
channel the correction is determined by the leptonic decay constant of $%
J/\Psi$. These corrections are so large that the predictions for the
considered process may be unreliable, if one only keeps several leading
terms in the small velocity expansion. An
integrated spin observable is proposed to measure the $J/\Psi$ polarization.
If detailed information of the eight matrix elements is known, numerical
values of polarized decay widths and the observable may be obtained.

\vskip40pt

\vskip20pt \noindent
{\bf Acknowledgment:} The author would like to thank Prof. K.T. Chao and
Prof. Y.Q. Chen for discussions. This work is supported by National Science
Foundation of P.R. China and by the Hundred Yonng Scientist Program of
Sinica Academia of P.R.China.

\vskip15pt

\vfil\eject

\end{document}